\documentclass[12pt]{article} 
\usepackage{epsfig}  
\def\bea{\begin{eqnarray}}
\def\eea{\end{eqnarray}}
\def\beq{\begin{equation}}  
\def\eeq{\end{equation}}  
  
\def\beqn{\begin{eqnarray}}  
\def\eeqn{\end{eqnarray}}  
\def\l{\left}
\def\r{\right}
\def\ra{\rightarrow}
  
\relax

\def\l{\left}  
\def\rhs{RHS~}

\def\pl#1#2#3{{\it Phys. Lett. } {\bf #1} (19#2) #3}

\def\pr#1#2#3{{\it Phys. Rev. } {\bf #1} (19#2) #3}  
\def\np#1#2#3{{\it Nucl. Phys. } {\bf #1} (#2) #3}  
\def\epj#1#2#3{{\it Eur. Phys. J.} {\bf #1} (#2) #3}

\def    \hepph  #1 {{\tt hep-ph/#1}}  
\def    \hepex  #1 {{\tt hep-ex/#1}}  

\newcommand\sss{\scriptscriptstyle}  
  
\newcommand\as{\alpha_{\sss S}}

\newsavebox\tmpfig  
  
\begin{document}

\pagestyle{empty}  
  
\begin{flushright}   
  
{\tt hep-ph/0107108}\\GEF/TH-11-01\end{flushright}   
  
\begin{center}   
\vspace*{0.5cm}  
{\Large \bf Solving the Altarelli-Parisi equations with truncated moments}\\  
\vspace*{1.5cm}   
{\bf Andrea Piccione} \\
\vspace{0.6cm}  
{}Dipartimento di Fisica, Universit\`a di Genova \\and\\ {}INFN,  
Sezione di Genova,\\  
Via Dodecaneso 33, I-16146 Genova, Italy\\  
\vspace*{1.5cm}  
  
{\bf Abstract}

\end{center}

\noindent  

The technique of truncated moments of parton distributions allows us
to study scaling violations without making any assumption on the shape
of parton distributions. The numerical implementation of the method is
however difficult, since the evolution equations for truncated moments
are not diagonal. We present a simple way to improve the efficiency of
the numerical solution of the evolution equations for truncated
moments. As a result, the number of truncated moments needed to
achieve the required precision in the evolution is significantly
smaller than in the original formulation of the technique.  The method
presented here can also be used to obtain the value of parton
distributions in terms of truncated moments, and therefore it can be
viewed as a technique for the solution of the Altarelli-Parisi
equations.

\vspace*{1cm}

\vfill  
\noindent  
  
\begin{flushleft} July 2001 \end{flushleft}   
\eject   
  
\setcounter{page}{1} \pagestyle{plain}  

The measurement of deep-inelastic scattering structure functions is a
central issue in strong interaction physics: it allows the extraction
of the parton distributions of hadrons, which, though in principle
computable, are determined by the nonperturbative dynamics of the
theory, and must be treated as unknown phenomenological parameters. A
detailed understanding of these quantities is an essential ingredient
of phenomenology at hadron colliders~\cite{qcd,pdfrev}.  Furthermore,
the measurement of scaling violations allows a determination of the
only free parameters in the QCD lagrangian, the strong coupling $\as$,
and of the gluon density in the nucleon.

As is well known, scaling violations of parton distribution functions
are described by the Altarelli-Parisi (AP) evolution
equations~\cite{AP}. These are integro-differential equations whose
kernels are presently known to next-to-leading order~\cite{kis} (for a
handful of operators, the anomalous dimensions are known to
next-to-next-to-leading accuracy~\cite{verm}). The coefficient
functions that determine the relation between measured structure
functions and parton distributions have also been computed up to
next-to-next-to-leading order~\cite{willy}.

There are different techniques to solve the Altarelli-Parisi
equations, and the choice among them depends on the kind of
problem at hand. Usually we solve the evolution equations analytically
by taking their Mellin transform, which turns convolution products 
into ordinary ones, and therefore the $x$-space integro-differential
equation into a set of independent ordinary first order differential
equations.  A parametrization of the distributions is assigned at some
initial scale, and the parameters are then determined by fitting to
data the evolved distributions. Mellin moments of structure functions,
however, cannot be measured even indirectly, since they are defined as
integrals over the whole range $0\leq x\leq 1$, and thus require
knowledge of the structure functions for arbitrarily small $x$, {\it
i.e.} arbitrarily large energy.

We can solve this problem using the Altarelli-Parisi equation
to evolve parton distributions directly: the scale dependence of any parton  
distribution at $x_0$ is then determined by knowledge of parton  
distributions for all $x > x_0$, {\it i.e.}, parton evolution is  
causal. In fact, through a judicious choice of factorization  
scheme~\cite{aem,cata} all parton distributions can be identified with  
physical observables, and it is then possible to use the  
Altarelli-Parisi equations to express the scaling violations of  
structure functions entirely in terms of physically observable  
quantities. It is, however, hard to measure local scaling violations  
of structure functions in all the relevant processes: in practice, a  
detailed comparison with the data requires the solution of the  
evolution equations.

As pointed out, the solution of the evolution equations 
requires an assumption on the
$x$ dependence of the parton distributions at the initial scale; a
frequently-adopted input is for example~\cite{pdfparref} 
\beq
q(x,Q_0^2)=a_0\,x^{a_1}\,(1-x)^{a_2}\,P(x;,a_3,\ldots)\,,
\label{pdfpar}
\eeq 
where $Q_0^2$ is a reference scale.  The parameter $a_1$ is associated
with the small-$x$ behavior while $a_2$ is associated with the
large-$x$ valence counting rules.  The term $P(x;,a_3,\ldots)$ is a
suitably chosen smooth function, depending on one or more parameters,
that adds more flexibility to the parton distributions
parametrization. It has however become increasingly clear that in
practice this procedure introduces a potentially large theoretical
bias, whose size is very hard to assess~\cite{pdfrev}. In ref.~\cite{pdfer}
it was proposed to adopt a functional method to keep this theoretical
error under control. Another suitable way to minimize the bias introduced
by the parton distributions parametrization is to project the parton
distributions on an optimized basis of orthogonal functions. Different methods
have been suggested with suitable families of orthogonal polynomials
({\it e.g.}~Bernstein \cite{ynd}, Jacobi \cite{par} or Laguerre polynomials
\cite{fur}) as basis of function.

A different approach has been suggested in refs.~\cite{FM,fmpr}, which
makes use of truncated moments of parton distributions.  Truncated
moments are defined in analogy with ordinary moments, but the
integration in the Bjorken variable $x$ is now restricted to the
subset $x_0\le x\le 1$ of the allowed kinematic range $0\le x\le 1$.
As a consequence, the corresponding evolution equations are not in
diagonal form: the evolution of the moment of order $n_0$ depends on
all moments of order $n_0+k$, with $k>0$. The solution of the
evolution equations is more complicated than in the case of ordinary
moments, but it can be performed by taking only a finite set of
truncated moments. Assigning their values at a reference scale as
input parameters, we can fit the value of $\as$ or the first moment of
the gluon; the initial values of truncated moments can be obtained
directly by data. In this way, no assumptions are made on the shape of
parton distributions. The fit is only affected by the experimental
errors and the theoretical uncertainty that affects the $Q^2$ evolution
can be easily kept under control. However, it was shown in
ref.~\cite{fmpr} that the number $M$ of truncated moments needed to
achieve a precision on the evolution of the lowest moment
comparable to that of other techniques is rather
large ($M\sim 150$), and in some cases it may lead to problems in the
numerical implementation of the method.  This problem was overcome in
ref.~\cite{fmpr} by showing that, in practice, it is sufficient to
parametrize the parton distributions using the first few (between 7
and 10) truncated moments, plus the value of the parton distributions
at $x=x_0$.

In this paper we present a different way to improve the numerical
efficiency of the method of truncated moments.  We will find that
integrating by parts the RHS of the evolution equations for truncated
moments allows us to use a significantly smaller number of truncated
moments involved in the evolution ($M\sim 10$).  We will see that the
integration by parts introduces an explicit dependence on $q(x_0,Q^2)$
of the AP equations, which complicates the solution of the evolution
equations for truncated moments. We will show how to circumvent this
difficulty. As a by-product, a formula is derived for the evolved
parton distribution at all values of $x$ larger than $x_0$, in terms
of the first $M$ truncated moments; the method provides therefore
an alternative way of solving the evolution equations.

We begin by studying the unpolarized non-singlet case at leading order, 
leaving at the end the extension to next-to-leading order.
The $Q^2$ dependence of parton distributions $q(x, Q^2)$ is governed  
by the Altarelli-Parisi equations~\cite{AP}  
\beq  
\frac{d}{dt}~q(x, Q^2)= \frac{\as(Q^2)}{2\pi}\int_x^1 \frac{d y}{y}   
P\left(\frac{x}{y}; \as (Q^2)\right) q(y, Q^2)~,  
\label{alpar}  
\eeq  
where $t = \log Q^2/\Lambda^2$. The evolution kernels $P(x,\as(Q^2))$ are
perturbatively computable as power series in $\as$. In the
non-singlet case, $q(x, Q^2)$ is simply one of the flavor non-singlet
combinations of quark distributions and $P(x,\as(Q^2))$ the corresponding
splitting function.

The truncated moments of a generic function $f(x)$ are defined as
\beq   
f_n(x_0) = \int_{x_0}^1 dx x^{n-1} f(x)\,.    
\eeq   
The evolution equations for truncated moments of
parton distributions were derived in
\cite{FM,fmpr}. One finds that the truncated moments of
$q(x,Q^2)$ obey the equation
\beq
\frac{d}{d\tau}q_n(x_0,Q^2) = \int_{x_0}^1 dy \,y^{n-1}
q(y,Q^2) G_n \l(\frac{x_0}{y}\r)
\label{eqev}
\eeq
with
\beq
\label{Gndef}
G_n (x) = \int_x^1 dz \,z^{n-1} P(z)
\eeq
and
\beq  
\tau = \int_{t_0}^t dt' \, a(t') \, ;   
\;\;\;\; a(t) = \frac{\as(Q^2)}{2\pi}\,.  
\label{tau}  
\eeq  
In \cite{FM,fmpr} it was shown that eq.~(\ref{eqev}) can be written as
\beq
\frac{d}{d\tau}q_n(x_0,Q^2) = \sum_{l=n_0}^{M+n_0} C_{nl} q_{l} (x_0,Q^2)\,,
\label{paper}
\eeq
where $C_{nl}$ are the elements of a triangular matrix, and only a
finite number $M$ of truncated moments is taken into account.  It was
also shown that in order to reach a precision of $5\%$ on the \rhs of
the evolution equation for the lowest value of $n_0$ a large value of
$M$ is needed: $M\sim 150$. This makes it difficult to solve the
evolution equation as we need a large numerical precision. In the
following we will show how these difficulties can be overcome. For
later convenience we will set $n_0=1$.

We now integrate by parts the RHS of eq.~(\ref{eqev}). We get
\beq
\label{rhs}
\int_{x_0}^1 dy\,y^{n-1} q(y,Q^2) G_n \l(\frac{x_0}{y}\r) =
\l[\widetilde{G}_n (x_0,y) y^{n-1} q(y,Q^2)\r]_{x_0}^1 -
\int_{x_0}^1 dy \,\widetilde{G}_n (x_0,y) \frac{d}{dy}
\l(y^{n-1} q(y,Q^2)\r) \nonumber
\eeq
where 
\beq
\widetilde{G}_n (x_0,y) = \int_{x_0}^y \,dz\,G_n \l(\frac{x_0}{z}\r)
\label{gt1}
\eeq
(the lower integration bound is irrelevant here; it has been chosen
equal to $x_0$ for later convenience). Using the definition of $\widetilde{G}_n
(x_0,y)$ and eq.~(\ref{Gndef}), we get 
\beq
\widetilde{G}_n (x_0,y)
=\int_{x_0}^y dz\int_{x_0/z}^1dx\,x^{n-1}P(x)
=\int_{x_0/y}^1dx\,x^{n-1}P(x)\int_{x_0/x}^y dz
=yG_n \l(\frac{x_0}{y}\r) - x_0G_{n-1} \l(\frac{x_0}{y}\r)\,.
\eeq
By taking the Taylor expansion of $\widetilde{G}_n (x_0,y)$ 
around $y=1$, we obtain 
\bea
\frac{d}{d\tau}q_n(x_0,Q^2) =
\l[\widetilde{G}_n (x_0,y) y^{n-1} q(y,Q^2)\r]_{x_0}^1
- \sum_{p=0}^{\infty}\frac{\widetilde{g}_n^p (x_0)}{p!}
\int_{x_0}^1 dy (y-1)^p \frac{d}{dy} \l(y^{n-1} q(y,Q^2)\r)
\label{tayexp1}
\eea
where
\bea
\widetilde{g}_n^p (x_0) = 
\l[\frac{d^p}{dy^p} \widetilde{G}_n (x_0,y)\r]_{y=1} = 
\l[\frac{d^{p-1}}{dy^{p-1}} \frac{d}{dy}\widetilde{G}_n
(x_0,y)\r]_{y=1} = g_n^{p-1} (x_0)\,.
\label{dertaycoeff}
\eea
The functions $\widetilde{G}_n(x_0,y)$ are regular in the whole
interval $[x_0,1]$. In fact, the $G_n(x_0/y)$ are regular for all
values of $y$ except $y=x_0$, as they contain singular terms
proportional to $\log(1-x_0/y)$. However, these terms are integrable,
and independent of $n$. Thus, $\widetilde{G}_n (x_0,y)$ is regular in
the limit $y\ra x_0$  and tends to zero.  Furthermore, we observe
that the Taylor coefficient of order $p$ of $\widetilde{G}_n(x_0,y)$
is equal to that of $G_n(x_0/y)$, times a factor $1/p$ (see
eq.~(\ref{dertaycoeff})). For this reason, the convergence of the
expansion of $\widetilde{G}_n(x_0,y)$ is faster than that of
$G_n(x_0/y)$.

Integrating by parts the second term of the RHS of eq.~(\ref{tayexp1}), 
we have:
\bea
\frac{d}{d\tau}q_n(x_0,Q^2) &=&
\l[\widetilde{G}_n (x_0,y) y^{n-1} q(y,Q^2)\r]_{x_0}^1
- \l[\sum_{p=0}^{\infty}\frac{\widetilde{g}_n^p (x_0)}{p!} 
(y-1)^p y^{n-1} q(y,Q^2)\r]_{x_0}^1 \nonumber \\
&+& \sum_{p=1}^{\infty} \frac{g_n^{p-1} (x_0)}{(p-1)!}
\int_{x_0}^1 dy\,y^{n-1} (y-1)^{p-1} \,q(y,Q^2)
\eea

\noindent Truncating the series, expanding the binomial $(y-1)^{p-1}$
and imposing $q(1,Q^2)=0$ (this is our only assumption on the behavior 
of the parton distributions), we get
\bea
\frac{d}{d\tau}q_n(x_0,Q^2) = 
x_0^{n-1} q(x_0,Q^2)\,\sum_{p=0}^{M} 
\frac{\widetilde{g}_n^p (x_0)}{p!} (x_0-1)^p 
+ \sum_{k=0}^{M-1} c_{nk}^{(M-1)}(x_0) q_{n+k} (x_0,Q^2)
\eea
where 
\bea  
c^{(M)}_{nk}(x_0) = \sum_{p=k}^M \frac{(-1)^{p+k}   
g_{p}^n (x_0)}{k! (p-k)!}\,.  
\eea 
Defining the triangular matrix 
\bea  
\left\{  
\begin{array}{cccc}  
C_{k l} & = & c_{k,l - k}^{(M - k + n)} & (l \geq k)~~, \\  
C_{k l} & = & 0 &(l < k)~~,  
\end{array}  
\right.  
\label{matr0}  
\eea
we can finally write the truncated evolution equation as
\bea
\frac{d}{d\tau}q_n(x_0,Q^2) = 
x_0^{n-1} q(x_0,Q^2)\,\sum_{p=0}^{M} 
\frac{\widetilde{g}_n^p (x_0)}{p!} (x_0-1)^p 
+ \sum_{l=1}^{M} C_{nl} q_{l} (x_0,Q^2)\,.
\label{eqev1}
\eea
Notice that the first term in the \rhs of eq.~(\ref{eqev1}) vanishes
in the limit $M\ra\infty$ and the original expression given in
refs.~\cite{FM,fmpr} is recovered. However, for finite values of $M$
this term must be taken into account (in a sense, it is the price we
have to pay for the better convergence of the expansion after the
integration by parts).  This term poses special problems because it
depends on the value of the parton distributions at $x=x_0$. In the
following we will show how to obtain an approximated expression of
$q(x_0,Q^2)$ in terms of a finite number $N$ (not necessarily equal to
$M$) of truncated moments.  The evolution equation (\ref{eqev1}) will
then be solved with a technique similar to that presented in
\cite{FM,fmpr}.

We begin by taking the Taylor expansion of $q(x,Q^2)$
around $x=y_0$:
\bea
q(x,Q^2) = \sum_{k=1}^{\infty} \eta_k (Q^2)(x-y_0)^{k-1}\,,
\label{pdftaylor}
\eea
The initial point of the expansion, $y_0$, must be carefully chosen.
Parton distributions pa\-ra\-me\-tri\-zed as in eq.~(\ref{pdfpar}) are
non-analytical in $x=1$ when the exponent $a_2$ is not an integer; and
even in that case, an essential singularity in $x=1$ is generated by
perturbative evolution.  One should therefore choose
$y_0\leq(1+x_0)/2$, so that the expansion (\ref{pdftaylor}) is
convergent everywhere in $[x_0,1)$.  The series will not be convergent
in $x=1$, no matter what $y_0$ is; however, the singularity in $x=1$
is integrable, and the term-by-term integration is allowed using the
Lebesgue definition of the integral (see ref.~\cite{fmpr} for details).
We have therefore
\bea
\label{qjx0}
q_j (x_0,Q^2) = \int_{x_0}^1\,dx\, x^{j-1} q(x,Q^2) 
= \sum_{k=1}^{\infty} \beta_{jk}(x_0,y_0)\,\eta_k(Q^2)\,,
\eea
where
\bea
\label{betadef}
\beta_{jk}(x_0,y_0) = \int_{x_0}^1\,dx\, x^{j-1} (x-y_0)^{k-1}\,.
\eea
Our task is now to find a way of inverting eq.~(\ref{qjx0}), in order to
express the coefficients $\eta_k(Q^2)$ in terms of the truncated moments
$q_j (x_0,Q^2)$. This can be done in the following way.
Define a matrix $\widetilde{\beta}^{-1}$ by
\bea
\widetilde{\beta}_{kj}^{-1} = 
\left\{  
\begin{array}{cc}  
\l(\beta^{(N)}\r)^{-1}_{kj} & k,j \leq N \\
0 & \rm{otherwise}
\end{array}  
\right.  
\eea
where $\beta^{(N)}$ is the $N\times N$ upper left 
submatrix of $\beta$. For example, in the case $N=2$ the matrix
$\widetilde{\beta}^{-1}$ is such that
\bea
\widetilde{\beta}^{-1} \cdot \beta
= \l(\matrix{1 & 0 & \frac{\beta_{13}\beta_{22}-\beta_{23}\beta_{12}}{\det\beta^{(2)}} & \dots \cr
             0 & 1 & \frac{\beta_{23}\beta_{11} - \beta_{13}\beta_{21}}{\det\beta^{(2)}} & \dots \cr
             0 & 0 & 0                                     & \dots \cr
             \vdots     & \vdots & \vdots                  & \ddots
 }\r)
\eea
Multiplying eq.~(\ref{qjx0}) by $\widetilde{\beta}^{-1}$ on the right,
we obtain
\bea
\sum_{j=1}^{N} \widetilde{\beta}_{ij}^{-1}\,q_j (x_0,Q^2) =
\sum_{k=1}^{\infty} \sum_{j=1}^{N} \widetilde{\beta}_{ij}^{-1}\,
\beta_{jk}\,\eta_k(Q^2)\,,
\eea
where for simplicity we have not shown the dependence on $x_0,y_0$.
Using the definition of $\widetilde{\beta}^{-1}$ we get
\bea
\sum_{j=1}^{N} \widetilde{\beta}_{ij}^{-1}\,q_j (x_0,Q^2) \equiv
\eta_i(Q^2) +
\sum_{k=N+1}^{\infty} \sum_{j=1}^{N} \widetilde{\beta}_{ij}^{-1}\,
\beta_{jk}\,\eta_k(Q^2) 
\label{etadef}
\eea
for $i\leq N$.
Substituting eq.~(\ref{etadef}) in
eq.~(\ref{pdftaylor}) gives
\bea
q(x_0,Q^2) =
\sum_{k=1}^{N} \sum_{j=1}^{N} \widetilde{\beta}^{-1}_{kj}\,
q_j (x_0,Q^2)\, (x_0-y_0)^{k-1} +R(x_0,y_0,Q^2)\,,
\label{pdfapprox}
\eea
where
\bea
\label{errapprox}
R(x_0,y_0,Q^2)=\sum_{k=N+1}^{\infty} \eta_k(Q^2) \l[(x_0-y_0)^{k-1}-
\sum_{i=1}^{N} (x_0-y_0)^{i-1}\sum_{j=1}^{N} 
\widetilde{\beta}^{-1}_{ij}\,
\beta_{jk} \r] \,.
\eea
We have thus obtained an approximate expression of $q(x_0,Q^2)$ as a
function of the first $N$ truncated moments of $q$,
eq.~(\ref{pdfapprox}); the quantity $R$ in eq.~(\ref{errapprox})
represents the error on this reconstruction. The quantity in square
brackets in eq.~(\ref{errapprox}) is independent of the parton
distributions, and can be computed for any $N$ and $k$ starting from
the coefficients $\beta_{ij}$, given by eq.~(\ref{betadef}). 
The analytic expression of this quantity is very complicated. We have
checked that, for $y_0=(1+x_0)/2$, it decreases as
$[(x_0-1)/2]^{k-1}$, for any value of $N$. Therefore,
$R(x_0,y_0,Q^2)$ vanishes, for $N\to\infty$, at least as fast as the
remainder of order $N$ of the Taylor expansion in
eq.~(\ref{pdftaylor}).

In order to assess the accuracy of our approximation, we have
computed the percentage error given by ratio 
$|R/q(x_0,Q^2)|$ for some representative choices of
the parton density, namely $q(x,Q^2)=(1-x)^{a_2}$ with
$a_2=2.5,3.5,4.5$.  We have fixed $x_0=0.1$ and $y_0=(1+x_0)/2$.  The
results are shown in Table~\ref{tabpdf}. We see that an excellent
approximation is achieved already with $N=5$, independently of the
value of the large-$x$ exponent $a_2$.  The accuracy increases with
increasing $N$; however, it should be noted that a numerical
evaluation of the matrix $\widetilde{\beta}^{-1}$ requires a numerical
precision which also rapidly increases with $N$. Therefore, for a
practical implementation of the method, $N$ cannot be very large. We
see from Table~\ref{tabpdf} that for $5\leq N\leq 10$ the accuracy is
already better than $10^{-3}$ in the cases we have studied.
\begin{table}[htm]  
\begin{center}  
\begin{tabular}{|c||c|c|c|} 
\hline
\multicolumn{4}{|c|}{$x_0=0.1$}\\ \hline  
\hline  
N & $a_2=2.5$ & $a_2=3.5$ & $a_2=4.5$ \\
\hline
5  & $3.3\times 10^{-4}$ & $3.2\times 10^{-4}$ & $1.4\times 10^{-3}~$ \\ 
\hline
10 & $3.8\times 10^{-6}$ & $5.4\times 10^{-7}$ & $1.4\times 10^{-7}~$ \\ 
\hline
15 & $3.2\times 10^{-7}$ & $1.8\times 10^{-8}$ & $1.8\times 10^{-9}$ \\
\hline
\end{tabular}
  \caption{\it Precision in the reconstruction of
$q(x_0,Q^2)=(1-x_0)^{a_2}$ in terms of a finite number $N$ of
truncated moments, for different values of $N$ and for three different
choices of $a_2$.}
\label{tabpdf}
\end{center}
\end{table}
We conclude that
\bea
q(x_0,Q^2) \simeq \sum_{j=1}^{N} \l[\sum_{k=1}^{N} 
\widetilde{\beta}^{-1}_{kj}(x_0,y_0)\,(x_0-y_0)^{k-1}\r]q_j (x_0,Q^2)
\label{reconpdf}
\eea
to an accuracy of about $10^{-3}$ for $N=5$, 
independent of the detailed shape of $q(x,Q^2)$, and rapidly increasing
with $N$.

We are now ready to re-write the original evolution equation
(\ref{eqev1}) using our result eq.~(\ref{reconpdf}). We get
\bea
\frac{d}{d\tau}q_n(x_0,Q^2) =
 \sum_{l=1}^{M} C_{nl}\,q_{l} (x_0,Q^2)
+\sum_{l=1}^{M} D_{nl}^{(N)}\,q_{l} (x_0,Q^2)\,,
\label{eqev2}
\eea
where $C_{nl}$ is defined in eq.~(\ref{matr0}), and
\bea
D_{nl}^{(N)} =x_0^{n-1} 
\l[\sum_{p=0}^{M} \frac{\widetilde{g}_n^p (x_0)}{p!} (x_0-1)^p\r]\,   
\l[\sum_{k=1}^N \widetilde{\beta}^{-1}_{kl}(x_0,y_0)\,(x_0-y_0)^{k-1}\r]\,.
\label{Dnl}  
\eea

We now turn to a test of the accuracy of the evolution equation. We will also
compare the accuracy achieved with the method presented here, and that of
refs.~\cite{FM,fmpr}. The original evolution equation (\ref{eqev})
and its truncated version, eq.~(\ref{paper}), can be
written schematically as
\bea
\frac{d}{d\tau}q_n(x_0,Q^2) = S_n \qquad \qquad \qquad ~~
\frac{d}{d\tau}q_n(x_0,Q^2) = S_n^{(M)}
\eea
respectively. Therefore, the quantity
\bea
{\cal R}_{n,M}^a = 1-\frac{S_n^{(M)}}{S_n}
\eea
is a measure of the accuracy of the method adopted in refs.~\cite{FM,fmpr}. 
Similarly, we write eqs.~(\ref{eqev2}) in the form
\bea
\frac{d}{d\tau}q_n(x_0,Q^2) = S_n^{(M-1)} + T_n^{(M,N)}\,,
\eea
and define
\bea
{\cal R}_{n,M,N}^b=1-\frac{S_n^{(M-1)}+T_n^{(M,N)}}{S_n}
\eea
to test the error of the method presented above.
The values of ${\cal R}_{n,M}^{a,b}$, computed
at leading order with $x_0=0.1$ for $n=1$ and $n=2$,
are shown in Table~\ref{tabrhs} for
different values of $M$ and $N=6$.
\begin{table}[t]  
\begin{center}  
\begin{tabular}{|r||r|r||r|r|} \hline 
\multicolumn{5}{|c|}{$x_0=0.1$}\\ \hline  \hline  
$M$  & ${\cal R}_{1,M}^a$ & ${\cal R}_{1,M,6}^b$
&  ${\cal R}_{2,M}^a$ & ${\cal R}_{2,M,6}^b$ \\
\hline
5  & 0.62 & 0.14 & 0.14 & 0.020 \\
\hline
10 & 0.48 & 0.07 & 0.12 & 0.016  \\
\hline
20 & 0.33 & 0.03 & 0.09 & 0.009  \\
\hline
40 & 0.20 & 0.01 & 0.05 & 0.004  \\
\hline
\end{tabular}
\caption{\it Comparison between percentage errors for the first and the second 
truncated moment at LO with $N=6$, $x_0=0.1$, $y_0=(1+x_0)/2$
and $q(x,Q^2)=(1-x)^{3.5}$.}
\label{tabrhs}
\end{center}
\end{table}
We observe that the error ${\cal R}_{n,M,N}^b$ computed
with the technique presented here is always much smaller than
the corresponding error of refs.~\cite{fmpr}, ${\cal R}_{n,M}^a$.
An accuracy of less than $10\%$ can be achieved with a relatively small
value of $M$.

The complete solution of the evolution equation, LO and NLO terms, 
can be written as~\cite{fmpr}
\bea  
\label{q1ter}  
q_n(x_0,Q^2)=R^{-1}\left[e^{\gamma\tau}  
+ae^{(\gamma+b_0)\tau}\int_0^\tau d\sigma\, e^{-(\gamma+b_0)\sigma}  
(\widehat{C}_1+\widehat{D}_1) e^{\gamma\sigma}\right]R\,q_n(x_0,Q_0^2)\,,  
\eea  
with the initial condition  
\bea  
q_n(x_0,Q_0^2)=q_n^{(0)}(x_0,Q_0^2)+a(0)q_n^{(1)}(x_0,Q_0^2)\,,
\eea  
and
\bea  
R(C_0+D_0)R^{-1}&=&{\rm diag}(\gamma_1,\ldots,\gamma_M)\equiv \gamma\,,
\nonumber \\ \nonumber
\widehat{C}_1+\widehat{D}_1&=&R(C_1+D_1)R^{-1}\,. 
\eea
The matrix $R$ that diagonalizes $C_0+D_0$ must be computed numerically.
This is not a problem, since, as we have seen, its dimension $M$ does 
not need to be too large.

In conclusion, we have shown that the evolution of truncated moments
can be computed, to a degree of accuracy sufficient for practical
purposes, by solving a system of a reasonably small ($\sim 10$) number
of coupled differential equations. The improvement of the numerical
efficiency presented here can be straightforwardly extended to the
unpolarized singlet case, and to polarized partons as well.
The tests we presented are only for the LO equations. We have
checked that the inclusion of NLO terms does not modify our
conclusions. The technique of truncated moments can now be easily 
implemented numerically for phenomenological purposes.

We have also shown that truncated moments provide, through
eq.~(\ref{reconpdf}), a parametrization of the parton distribution
itself, which has the advantage of being free of theoretical biases
on the shape of the distribution at a given scale.

\vskip 1cm  
  
\noindent {\large {\bf Acknowledgements}}  

\vskip 0.5cm

I am greatly indebted to Giovanni Ridolfi for many useful discussion
and suggestions. I thank Stefano Forte and Lorenzo Magnea for
carefully reading the manuscript and several useful comments.

\end{document}